\documentclass[11pt]{article}
\usepackage{amsthm}
\usepackage{epsfig}
\usepackage{caption}
\usepackage{algorithm}
\usepackage{algorithmic}
\usepackage{latexsym}
\usepackage{amsmath}
\usepackage{amsfonts}
\usepackage{graphicx}
\usepackage[normalem]{ulem}
\usepackage{array}
\usepackage{enumerate}
\usepackage{multirow}
\usepackage{color}\usepackage[dvipsnames]{xcolor}
\usepackage{tikz}\usetikzlibrary{patterns}
\usepackage[
  bookmarks=false, 
  colorlinks,
  citecolor=brown!70!black,
  linkcolor=brown!80!black,
  urlcolor=blue!70!black,
]{hyperref}
\usepackage{relsize,exscale}

\newtheorem{lem}{Lemma}
\newtheorem{cor}{Corollary}
\newtheorem{thm}{Theorem}
\newtheorem{prop}{Proposition}

\theoremstyle{definition}

\theoremstyle{remark}
\newtheorem{fact}{Fact}

\newtheorem{exam}{Example}

\newcommand{\comm}[1]{}

\newcommand{\commm}[1]{#1}

\title{Popularity of patterns over $d$-equivalence classes of words and permutations}
\author{Jean-Luc {Baril} and Vincent Vajnovszki\\  
LIB, Universit\'e de Bourgogne Franche-Comt\'e\\
         B.P. 47 870, 21078 Dijon-Cedex France      \\
        {\tt e-mail:\{barjl,vvajnov\}@u-bourgogne.fr}}

\begin{document}
\maketitle

\begin{abstract} 
Two same length words are $d$-equivalent if they have same descent set
and same underlying alphabet. In particular, two same length permutations are $d$-equivalent
if they have same descent set.
The popularity of a pattern in a set of words is the overall number of copies of the pattern within 
the words of the set.
We show the far-from-trivial fact that two patterns are $d$-equivalent if and only if they 
are equipopular over any $d$-equivalence class,
and this equipopularity does not follow obviously from a trivial equidistribution.
\end{abstract}

\section{Introduction and notation}

We consider words over the set of positive integers and permutations are
particular words. For $q\geq 1$, $[q]$ denotes the alphabet $\{1,2,\dots,q\}$ and
the {\it underlying alphabet} of a word is the set of symbols occurring in the word.
For instance, the $4$-ary words $4313$ and $4212$ have different underlying alphabet, namely
$\{1,3,4\}$ and $\{1,2,4\}$.
A {\it descent} in a word $w_1w_2\dots w_n$ is a position $i$ with 
$w_i>w_{i+1}$ and the {\it descent set}
of $w$, denoted $\mathrm{Des}(w)$, is the set of all such $i$;
{\it ascent} and {\it ascent set}, $\mathrm{Asc}(w)$, are defined similarly.
Two same length words are {\it $d$-equivalent} if they have the same descent set
and the same  underlying alphabet. 
For instance the words $31443$ and $21332$ have the same descent set but not the same 
underlying alphabet (thus they are not $d$-equivalent), whereas $31443$ and $41131$ are $d$-equivalent
(the common underlying alphabet is $\{1,3,4\}$).
If $v$ and $w$ are $d$-equivalent and $v$ is a permutation, it follows that so is $w$.
A {\it $d$-equivalence class} is a maximal set of $d$-equivalent words.
For a word $w_1w_2\dots w_n$ the {\it reverse} $\mathrm{r}(w)$
of $w$ is the word $w=w_nw_{n-1}\dots w_1$; and
the {\it complement} $\mathrm{c}(w)$ of $w$ is the word $(q-w_1+1)(q-w_2+1)\dots (q-w_n+1)$ with $q$
the maximal entry of $w$.

A {\it pattern} is a word with the property that if $i$ occurs in it, then so does $j$, for any $j$ with $1\leq j\leq i$, and the {\it reduction} of $w$, $red(w)$, is the unique pattern order isomorphic with $w$.
The pattern $\pi=\pi_1\pi_2\dots \pi_k$ occurs in the word $w=w_1w_2\dots w_n$, $k\leq n$,
if $w$ has a subword $w_{i_1}w_{i_2}\dots w_{i_k}$ order isomorphic with $\pi$,
see for instance Kitaev's seminal book \cite{Kitaev2011} on this topic.
The number of occurrences of the pattern $\pi$ in the word $w$ is denoted by
$(\pi)w$. 
For a set of words $S$, $(\pi)w$ becomes an integer valued statistic on $S$ and 
the overall number of occurrences of $\pi$ within the words of
$S$ is called the
{\it popularity} of $\pi$ in $S$; more formally, the popularity of $\pi$ in $S$ is 
$\sum_{w\in S}(\pi)w$.

The equidistribution of two patterns implies their equipopularity and 
recently a growing interest is shown in patterns that have same popularity 
but not same distribution on particular classes of words or permutations,
see for instance \cite{AHP2015,Bona2012,Homberger,Rudo}.

In this paper we show that two patterns are $d$-equivalent if and only if they have the same popularity
on any $d$-equivalence class. Specializing to permutations, we obtain that two same length permutations 
have same descent set  
if and only if they have the same popularity on any descent-set equivalence class of permutations.

\section{Preliminary notions and results}

\subsection{$d$-equivalence vs. $f$-equivalence}

Here we show that any two $d$-equivalent patterns can be obtained from each other by a sequence
of $f$-transformations, which are `small changes' preserving the $d$-equivalence class.

\subsubsection*{Lexicographically smallest pattern in a $d$-equivalence class}

For our purposes we need the lexicographically minimal pattern $d$-equivalent with
a given pattern $\pi$,
which in turn requires two other particular patterns $\alpha$ and $\omega$ that we define below.

The {\it descent word} of a length $n$ word $w$ is the binary word $b=b_1b_2\ldots b_n$ 
where $b_i=1$ if and only if $i$ is a descent in $w$ (and so, $b_n$ is a redundant $0$).
The minimal arity of a pattern having descent word $b$ is one more than the maximal number of consecutive $1$s in $b$,
and we denote by $\alpha(b)$ the lexicographically smallest pattern having minimal arity and descent word $b$.
It is easy to see that the pattern $\alpha=\alpha(b)$ is defined as:
for each $i$, $1\leq i\leq n$, 
$$
\alpha_i=\min\{j:j\geq i,\ b_j=0\}-i+1.
$$

\begin{exam}
With $n=9$ and  $\pi=\mathbf{43}2\mathbf{4}112\mathbf{3}1$, we have  $b=110 10 00 10$ and  
$\alpha=\mathbf{32}1 \mathbf{2}111\mathbf{2}1$ (the descents are in bold).
\end{exam}

The maximal arity of a pattern having descent word $b=b_1b_2\dots b_n$ is $n$,
and we denote by $\omega(b)$ the lexicographically smallest $n$-ary pattern having descent word $b$,
which is necessarily a length $n$ permutation.
We divide the descent word $b$ into {\it runs}: the length maximal factors
of the form $11\cdots 10$ with at least one occurrence of $1$ are {\it descent runs} and, for convenience,
we call the remaining length maximal $0$s factors (if any) {\it ascent runs}.
We define an order relation on $\{1,2,\ldots,n\}$:
for two integers $i$ and $j$, $1\leq i,j\leq n$, we say that $i$ {\it precedes} $j$, 
with respect to the binary word $b$, if
\begin{itemize}
\item $b_i$ and $b_j$ are in two distinct runs in $b$, and $i<j$, or
\item $b_i$ and $b_j$ are in the same ascent run in $b$, and $i<j$, or 
\item $b_i$ and $b_j$ are in the same descent run in $b$, and $i>j$.
\end{itemize}
The desired permutation $\omega=\omega(b)$ is precisely that induced by this order relation:
$$
\omega_i=\mbox{the rank of }i\mbox{ in }\{1,2,\ldots,n\},\mbox{ in the precedence order,} 
$$
and $\omega$ is at the same time the lexicographically minimal word of maximal
arity (that is $n$) having descent word $b$ and, as we will see below, defines an order 
in which we cover the entries of a pattern with descent word $b$.


\begin{exam} If $b$ is as in the previous example, then $b=110.10.00.10$ 
(runs are separated by dots) and $\omega=\mathbf{32}1\mathbf{5}467\mathbf{9}8$
(the descents are in bold).
\end{exam}

Now for an arbitrary arity $q\leq n$ (not necessarily its minimal, or its maximal value $n$), we construct
the lexicographically minimal pattern $\beta=\beta(q,b)$ where 
each symbol in $[q]$ occurs at least once and $i$ 
is a descent in $\beta$ if and only if $b_i=1$, and in 
this construction the above defined patterns $\alpha$ and $\omega$ are involved.
Moreover, if $q$ reaches its minimal value, then $\beta=\alpha$ and if 
$q=n$, then $\beta=\omega$.
The pattern $\beta$ is obtained by covering its entries in 
$\beta_{\omega_1},\beta_{\omega_2},\dots,\beta_{\omega_n}$
order and the first entries (in this order) are taken from $\alpha$ and the last ones are increasing integers
to guarantee that all symbols in $[q]$ occur in $\beta$.
Formally:

$$
\beta_{\omega_i}=\left\{ \begin {array}{ll}
\alpha_{\omega_i} & \mbox{if} \max\{\alpha_{\omega_1},\alpha_{\omega_2},\dots,\alpha_{\omega_i}\}\geq q-(n-i) \\
q-(n-i) & \mbox{elsewhere}.
\end {array}
\right.
$$
With these notations, it follows that if $\beta\neq \alpha$ and $k=\min\{i: \beta_{\omega_i}\neq \alpha_{\omega_i}\}$ then
$$
\beta_{\omega_i}=\left\{ \begin {array}{ll}
\alpha_{\omega_i} & \mbox{if } i<k \\
q-(n-i) & \mbox{elsewhere},
\end {array}
\right.
\label{def_k}
$$
and the entries $\beta_{\omega_k},\beta_{\omega_{k+1}},\dots,\beta_{\omega_{n}}$ are 
consecutive integers in increasing order.

\begin{exam}
Continuing the previous example with $b=110 10 00 10$, if $q=7$, then the above construction gives
$\beta=\mathbf{32121}4576$; 
and if $q=8$, then it gives $\beta=\mathbf{321}4\mathbf{1}5687$ (the entries taken from $\alpha$ are in bold).
\end{exam}

For a pattern $\pi$ with descent word $b$, by a slight abuse of notation we denote by
$\alpha(\pi)$ the pattern $\alpha(b)$, by $\omega(\pi)$ the pattern (permutation) $\omega(b)$;
and in addition if $\pi$ has arity $q$, then we denote
by $\beta(\pi)$ the pattern $\beta(q,b)$. 
Note that
\begin{itemize}
\item the pattern $\alpha(\pi)$ is lexicographically minimal in its $d$-equivalence class and so are 
      $\beta(\pi)$ and $\omega(\pi)$,
\item the four patterns $\pi$, $\alpha(\pi)$, $\beta(\pi)$ and $\omega(\pi)$ have the same
descent set, 
\item $\pi$ and $\beta(\pi)$ are $d$-equivalent, but
      $\pi$, $\alpha(\pi)$ and $\omega(\pi)$ are not necessarily $d$-equivalent since they can have different 
underlying alphabet (or equivalently in this case, different arity).
\end{itemize}

\subsubsection*{$f$-equivalent patterns}

For later use we need the following rather technical notion: for two $d$-equivalent patterns $\pi$ and $\sigma$ we say that $\sigma$ is an {\it $f$-transformation} of $\pi$
if $\sigma$ can be obtained from $\pi$ by 
either
\begin{itemize}
\item increasing or decreasing by $1$
an entry in $\pi$, or
\item interchanging in $\pi$ two entries with consecutive values.
\end{itemize}
Actually, the $f$-transformation is a symmetric binary relation on a set of $d$-equivalent patterns and
two patterns are said {\it $f$-equivalent} if they belong to the same equivalence class with respect
to the transitive closure of $f$-transformation.
Below we prove that the notions of $d$-equivalence and $f$-equivalence coincide, 
which is stated in Corollary \ref{main_cor_eq} of Theorem \ref{main_thm_eq}.

The order induced by $\omega(\pi)$ is related to the descent word of $\pi$, however we have the following.
\begin{prop}
Let $\pi$ be a pattern. If $\omega=\omega(\pi)$ and $i,k$ are such that $\pi_i>\pi_k$, then $\omega_i<\omega_k$ implies
$i<k$.
\end{prop}
\commm{
\proof
If $\pi_i>\pi_k$ and $\omega_i<\omega_k$, then $\omega_i$ and $\omega_k$ are not
in the same descent run of $\pi$, so $i<k$.
\endproof
}
%

The next proposition says that, under certain conditions, 
decreasing an entry in the pattern $\pi$ produces 
a $d$-equivalent pattern lexicographically smaller 
than $\pi$. 
\begin{prop}
Let $\pi$ be a pattern.
If $\omega=\omega(\pi)$, $\beta=\beta(\pi)$ and
\begin{itemize}
\item
there is an $i$ such that $\pi_{\omega_j}=\beta_{\omega_j}$ for any $j$, $1\leq j<i$,
\item
$\pi_{\omega_i}>\beta_{\omega_i}$,
\item 
the entry $\pi_{\omega_i}$ occurs at least twice in $\pi$,
\end{itemize}
then the word $\sigma$ with $\sigma_{\omega_j}=\pi_{\omega_j}$ for any $j$ except  
$\sigma_{\omega_i}=\pi_{\omega_i}-1$ is a pattern $d$-equivalent with $\pi$, lexicographically smaller than 
$\pi$.
\label{Pr2}
\end{prop}

\begin{prop}
Let $\pi$ be a length $n$ pattern.
If $\omega=\omega(\pi)$, $\beta=\beta(\pi)$ and
\begin{itemize}
\item
there is an $i$ such that $\pi_{\omega_j}=\beta_{\omega_j}$ for any $j$, $1\leq j<i$, 
\item
$\pi_{\omega_i}>\beta_{\omega_i}$, 
\item the entry $\pi_{\omega_i}$ occurs once in $\pi$, 
\item the entry $\pi_{\omega_i}-1$ occurs at least once in the set 
$\{\pi_{\omega_{i+1}},\pi_{\omega_{i+2}},\dots ,\pi_{\omega_n}\}$,
\end{itemize}
then there is a $k$ with $\omega_k\in\{\omega_{i+1},\omega_{i+2},\dots ,\omega_n\}$ and
$\pi_{\omega_k}=\pi_{\omega_i}-1$, and  
the word $\sigma$ with $\sigma_{\omega_j}=\pi_{\omega_j}$ for any $j$, except 
$\sigma_{\omega_i}=\pi_{\omega_i}-1$ and $\sigma_{\omega_k}=\pi_{\omega_k}+1\ (=\pi_{\omega_i})$
is a pattern $d$-equivalent with $\pi$, lexicographically smaller than 
$\pi$.
\label{Pr3}
\end{prop}
\proof
\commm{
Let $\omega_a$ be the largest element of the set $\{\omega_{i+1},\omega_{i+2},\dots ,\omega_n\}$
with $\pi_{\omega_a}=\pi_{\omega_i}-1$. It is enough to choose $k=a$. 
}
\endproof

Note that, in the two propositions above $\sigma$ is obtained by an $f$-transformation 
of~$\pi$.

\begin{prop} Let $\pi$ be a $q$-ary length $n$ pattern.
If $\omega=\omega(\pi)$,  $\beta=\beta(\pi)$ and
$i$ is such that each of $\beta_{\omega_i},\beta_{\omega_{i+1}},\dots, \beta_{\omega_n}$ occurs once in $\beta$,
then 
\begin{itemize}
\item[1.]
$\beta_{\omega_i},\beta_{\omega_{i+1}},\dots,\beta_{\omega_n}$ is a sequence of consecutive integers ending by 
$q$.
\end{itemize}
In addition, if $\pi_{\omega_j}=\beta_{\omega_j}$ for any $j$, $1\leq j<i$, then 
\begin{itemize}
\item[2.]
each of $\pi_{\omega_i},\pi_{\omega_{i+1}},\dots,\pi_{\omega_n}$ occurs once in $\pi$.
\end{itemize}
\label{Pr4}
\end{prop}
\proof
\commm{If condition 1. is violated, then $\beta$ is not the lexicographically smallest pattern in its
$d$-equivalence class, which is a contradiction. If condition 2. is violated, then
$\pi$ cannot be a $q$-ary pattern, again a contradiction.
}
\endproof

\begin{prop} 
Let $\pi$ be a length $n$ pattern. If $\omega=\omega(\pi)$, $\beta=\beta(\pi)$ and
\begin{itemize}
\item
there is an $i$ such that $\pi_{\omega_j}=\beta_{\omega_j}$ for any $j$, $1\leq j<i$, 
\item
$\pi_{\omega_i}>\beta_{\omega_i}$, 
\item the entry $\pi_{\omega_i}$ occurs once in $\pi$, 
\item the entry $\pi_{\omega_i}-1$ does not occur in the set 
$\{\pi_{\omega_{i+1}},\pi_{\omega_{i+2}},\dots ,\pi_{\omega_n}\}$,
\end{itemize}
then there is a pattern $\sigma$ which is $f$-equivalent with $\pi$ and lexicographically smaller than $\pi$.
\label{Pr5}
\end{prop}
\proof
\commm{
First we prove that at least one of the entries $\pi_{\omega_{i+1}},\pi_{\omega_{i+2}},\dots,\pi_{\omega_n}$
occurs at least twice in $\pi$. Indeed, if these entries occur once in $\pi$
so are the entries in the set 
$P=\{\pi_{\omega_i},\pi_{\omega_{i+1}},\pi_{\omega_{i+2}},\dots,\pi_{\omega_n}\}$.
But since $\pi$ and $\beta$ have the same arity and $\pi_{\omega_j}=\beta_{\omega_j}$ for any $j<i$
it follows that the two sets $P$ and 
$\{\beta_{\omega_i},\beta_{\omega_{i+1}},\dots,\beta_{\omega_n}\}$ are equal, and by the point 1.
of Proposition \ref{Pr4}, they are formed by consecutive integers. This is a contradiction
since $\pi_{\omega_i}$ is not the minimal element of $P$ 
(otherwise $\pi_{\omega_i}=\beta_{\omega_i}$) and $\pi_{\omega_i}-1$ does not occur in $P$.

Now we prove the statement according to the following two (non exclusive) cases: (i) there is an integer
larger than $\pi_{\omega_i}$ in the set $\{\pi_{\omega_{i+1}},\pi_{\omega_{i+2}},\dots,\pi_{\omega_n}\}$
that occurs at least twice in $\pi$, or (ii) there is at least one integer 
smaller than $\pi_{\omega_i}$ in the set $\{\pi_{\omega_{i+1}},\pi_{\omega_{i+2}},\dots,\pi_{\omega_n}\}$.

\noindent
Case (i).
If there is an integer larger than $\pi_{\omega_i}$ in the set $\{\pi_{\omega_{i+1}},\pi_{\omega_{i+2}},\dots,\pi_{\omega_n}\}$
that occurs twice in $\pi$, let $v\neq \pi_{\omega_i}$ be the smallest of them and 
let $\{\omega_{k_1},\omega_{k_2},\dots\}\subset
\{\omega_1,\omega_2,\dots,\omega_n\}$ with $k_1<k_2\cdots$ be the set of occurrences of $v$ in $\pi$.
The entry $v-1$ occurs once in $\pi$ and let define $a$ as: if $\pi_{\omega_{k_1}}$ is not in the same descent run as 
$v-1$, then
$a=\omega_{k_1}$, and $a=\omega_{k_2}$ otherwise. 
It follows that the pattern 
$\sigma$ with $\sigma_{\omega_j}=\pi_{\omega_j}$ for all $j$, except $\sigma_{\omega_a}=\pi_{\omega_a}-1$, is
lexicographically smaller than $\pi$ and is obtained from $\pi$ by an $f$-transformation. 

\noindent
Case (ii).
If there is an integer smaller than $\pi_{\omega_i}$ in the set $\{\pi_{\omega_{i+1}},\pi_{\omega_{i+2}},\dots,\pi_{\omega_n}\}$, then let $v\neq \pi_{\omega_i}$ be the largest of them and let
$\omega_a$ be the largest element of  $\{\omega_{i+1},\omega_{i+2},\dots,\omega_n\}$ with
$\pi_{\omega_a}=v$.
Necessarily $v$ occurs at least twice in $\pi$; otherwise, since $v+1$ does not occur in 
$\{\pi_{\omega_{i+1}},\pi_{\omega_{i+2}},\dots,\pi_{\omega_n}\}$, the pattern 
$\sigma$ with  with $\sigma_{\omega_j}=\pi_{\omega_j}$ for any $j$, except $\sigma_{\omega_a}=v+1$ and
$\sigma_{\omega_k}=v$ for an appropriate $k<i$ with $\pi_{\omega_k}=v+1$,
is $d$-equivalent with $\pi$ and lexicographically smaller than $\pi$,
which is in contradiction with $\pi_{\omega_j}=\beta_{\omega_j}$ for any $j<i$.

Since $v$ occurs at least twice in $\pi$ it follows that
the pattern $\tau$ with $\tau_{\omega_j}=\pi_{\omega_j}$ for any $j$, except $\tau_{\omega_a}=\pi_{\omega_a}+1$,
is $d$-equivalent with $\pi$ (and lexicographically larger than $\pi$) and the entry 
$\tau_{\omega_a}$ occurs at least twice in $\tau$. Now two subcases can occur:
$\tau_{\omega_a}=\pi_{\omega_i}-1\ (=\pi_{\omega_a}+1)$ or 
$\tau_{\omega_a}<\pi_{\omega_i}-1$.

\noindent
When $\tau_{\omega_a}=\pi_{\omega_i}-1$, since $\tau_{\omega_i}=\pi_{\omega_i}$ it follows that  
$\tau$ is a pattern satisfying Proposition \ref{Pr3}, and 
the pattern $\sigma$ with $\sigma_{\omega_j}=\tau_{\omega_j}$ for any $j$,
except $\sigma_{\omega_i}=\tau_{\omega_i}-1\ (=\pi_{\omega_i}-1)$ and 
$\sigma_{\omega_a}=\tau_{\omega_a}+1\ (=\pi_{\omega_i})$, is 
$d$-equivalent with 
$\tau$ (and thus with $\pi$)
 and is lexicographically smaller than $\pi$.
The patterns $\pi$ and $\sigma$ are $f$-equivalent, and the statement holds.

\noindent
When $\tau_{\omega_a}<\pi_{\omega_i}-1$, since the entry $\tau_{\omega_a}$ occurs twice in $\tau$
we can find as previously a pattern $\tau^{(2)}$ $d$-equivalent with $\tau$ (and so with $\pi$) where the entry 
$\tau_{\omega_a}+1$ occurs twice in $\tau^{(2)}$. Iterating this procedure, we obtain a sequence of $d$-equivalent patterns
$\tau=\tau^{(1)},\tau^{(2)},\dots,\tau^{(k)}$ with $\tau^{(k)}_{\omega_j}=\pi_{\omega_j}$ for all $j\leq i$
and $\tau^{(k)}_{\omega_{b}}=\pi_{\omega_{i}}-1$ for some $b>i$. As above, 
the pattern $\sigma$ with $\sigma_{\omega_j}=\tau^{(k)}_{\omega_j}$ for any $j$,
except $\sigma_{\omega_i}=\tau^{(k)}_{\omega_i}-1\ (=\pi_{\omega_i}-1)$ and 
$\sigma_{\omega_b}=\tau^{(k)}_{\omega_b}+1\ (=\pi_{\omega_i})$ is $d$-equivalent with 
$\tau^{(k)}$ (and thus with $\pi$)
and lexicographically smaller than $\pi$. Moreover, each $\tau^{(p)}$
is obtained from $\tau^{(p-1)}$ by an $f$-transformation and the statement holds.
}
\endproof

By Propositions \ref{Pr2}, \ref{Pr3} and \ref{Pr5} we have the following theorem.
\begin{thm}\label{main_thm_eq}
Any pattern $\pi$ is $f$-equivalent with $\beta(\pi)$.
\end{thm}
\proof
\commm{Let $\pi$ be a pattern with $\pi\neq \beta(\pi)$. Then $\pi$ is in one of the cases 
stated in Propositions \ref{Pr2}, \ref{Pr3} or \ref{Pr5}, and according to these 
propositions there exists a pattern $f$-equivalent (and thus $d$-equivalent) with $\pi$ and lexicographically smaller than $\pi$, and eventually $\pi$ is $f$-equivalent with the lexicographically
smallest pattern in its $d$-equivalence class, that is with $\beta(\pi)$.
\endproof
}

\begin{cor}\label{main_cor_eq}
Two patterns are $d$-equivalent if and only if they are $f$-equivalent.
\label{main_cor1}
\end{cor}
\proof
\commm{By definition, $f$-equivalence implies $d$-equivalence. Conversely, if two patterns $\pi$ and $\sigma$
are $d$-equivalent, then $\beta(\pi)=\beta(\sigma)$, and by the previous theorem $\pi$ is $f$-equivalent 
with $\beta(\pi)$ and $\sigma$ is $f$-equivalent 
with $\beta(\sigma)$. Finally $\pi$ and $\sigma$ are $f$-equivalent.
}
\endproof

\subsection{Bijection $\psi$}

In the following we need a bijection on $[q]^n$ onto itself,
that we denote by $\psi$, and satisfying:
\begin{enumerate}
\item[(a)] $\psi$ preserves the underlying alphabet,
\item[(b)] the number of occurrences of the largest entry is the same in 
$w$ and in $\psi(w)$, and the same holds for the smallest entry in $w$ and in $\psi(w)$,
\item[(c)] $\psi$ transforms descent set into ascent set, that is, for any word $w$
          $\mathrm{Des}\,w=\mathrm{Asc}\,\psi(w)$. 
\end{enumerate}

In particular when $w$ is a permutation,
the complement transformation $\mathrm{c}$ satisfies the three properties above,
which in general is not longer true for arbitrary words, and we propose a 
bijection $\psi$ which satisfies these properties for any words, not necessarily permutations.
Its construction is based on the bijection $\phi$ on words defined in \cite{Fu_Hu_Vaj_17},
which in turn is built on Foata and Sch\"utzenberger \cite{Foa_Schu_78} bijection $j$ on permutations. 
The bijection $\phi:[q]^n\to [q]^n$ in \cite{Fu_Hu_Vaj_17} satisfies for any word $w$:

\begin{enumerate}
\item[(i)] $\phi(w)$ is a rearrangement of the symbols of $w$,
\item[(ii)] $\mathrm{Des}\,w=\{n-i\,:\,i\in\mathrm{Des}\,\phi(w)\}$, and
\item[(iii)] $\mathrm{Ides}\,w=\mathrm{Ides}\,\phi(w)$.
\end{enumerate}
See \cite{Kit_Vaj_16} for the definition of the set valued statistic $\mathrm{Ides}$
that we will not use here and for 
a weaker version of $\phi$ satisfying only (ii) and (iii) above. 
Note that from (i) it follows that $\phi$ preserves the underlying alphabet.

Based on the properties (i) and (ii) of $\phi$ it is easy to check that $\psi:[q]^n\to [q]^n$ defined as
\begin{equation}
\psi=\mathrm{r}\circ \phi
\label{psi}
\end{equation}
satisfies the above desiderata (a)--(c).
Indeed, properties (a) and (b) follow from (i), and property (c) follows from (ii).
Property (iii)
is a deep and remarkable feature of $\phi$ (that we will not make use of it) and in some sense 
our bijection $\psi$ is over endowed.
For instance,
$\phi(1321)=3211$,
$\phi(1232)=3122$ and
$\phi(3321)=3213$  (see \cite{Fu_Hu_Vaj_17}), and thus  
$\psi(1321)=1123$,
$\psi(1232)=2213$ and
$\psi(3321)=3123$.
\label{sect2.3}

\subsection{Pattern trace and word substitution}

For a word  $w=w_1w_2\dots w_n$ and a set $S=\{i_1,i_2,\dots,i_p\}\subseteq \{1,2,\dots,n\}$ we denote by
$w_S$ the subword $w_{i_1}w_{i_2}\dots w_{i_p}$ of $w$.

Let $t=t_1t_2\ldots t_k$ be a length $k$ word over $[q]\cup\{\Box\}$, 
$q\geq 1$, and $I(t)$ be the set $\{\ell: 1\leq\ell\leq k,\ t_\ell\neq \Box\}$. 
We say that $t$ is {\it a trace} of the pattern $\pi=\pi_1\pi_2\dots \pi_k$
if $t_i$ and $t_j$ have the same relative order ($<$, $=$, or $>$) as $\pi_i$ and $\pi_j$ have
whenever $i,j\in I(t)$. Equivalently,  $t$ is a trace of $\pi$ if the words $t_{I(t)}$ and  $\pi_{I(t)}$
are order-isomorphic.
In particular, when $t$ does not contain $\Box$, then $red(t)=\pi$; and $t$ formed only by $\Box$'s 
is  a trace of any pattern. 
It can happen that $t$ is a trace of several patterns. For instance,
for two same length patterns $\pi$ and $\sigma$, if 
the trace $t$ of $\pi$ is such that 
$\pi_\ell\neq\sigma_\ell$ implies $t_\ell=\Box$, then $t$
is a trace of $\sigma$ as well.

With $t$ a trace of a pattern $\pi$ and $I(t)$ as above,
for a word $w=w_1w_2\dots w_n$ and a set $A\subset \{1,2,\ldots,n\}$ of positions in $w$ we say that 
$t$ is a {\it trace of $\pi$ in $w$ at $A$} if $t_{I(t)}=w_A$ (and so, $|I(t)|=|A|$), and
a trace $t$ of $\pi$ in $w$ at $A$ can be seen as a partial occurrence of the pattern $\pi$ in $w$
with $\Box$ playing the role of `wild' symbol.
It can happen that several occurrences of $\pi$ in a word $w$ have trace $t$ at $A$, and 
we denote by $(t,A,\pi)w$ the number of these occurrences, and thus 
$(t,A,\pi)$ becomes an integer valued statistic on words.

\begin{exam}
If $\pi=1332$ and $\sigma=2331$ are two patterns, then $t=\Box 44 \Box$ and $t'=\Box 55 \Box$ are traces of 
both $\pi$ and $\sigma$.
Furthermore, if $w=154543$, then
\begin{itemize}
\item $1443$ is an occurrence in $w$
of $\pi$ with trace $t=\Box 44 \Box$ at $A=\{3,5\}$, 
\item $1554$ and $1553$ are occurrences in $w$ of $\pi$ with trace $t'=\Box 55 \Box$ at $A=\{2,4\}$,
and $(t',A,\pi)w=2$.
\end{itemize}
\end{exam}
\medskip

See Table \ref{Appendix_T1} in Appendix for other examples.

For a word $w=w_1w_2\dots w_n$ and two pairs of integer $a<b$ and $c<d$ we denote by 
$w\,|\,([a,b],[c,d])$ the length-maximal subword $w_{i_1}w_{i_2}\dots w_{i_k}$ of $w$ with 
$\{i_1,i_2,\dots,i_k\}\subseteq[a,b]$ and  
$\{w_{i_1},w_{i_2},\dots, w_{i_k}\}\subseteq[c,d]$.
Alternatively, $w\,|\,([a,b],[c,d])$ is the length-maximal subword of $w_{[a,b]}$ with entries in  $[c,d]$.

If the word $w_{i_1}w_{i_2}\dots w_{i_k}=w\,|\,([a,b],[c,d])$ has $m$ different symbols
and $u=u_1u_2\dots u_k$ is a word with the underlying alphabet $[m]$, then there is a unique word $v=v_1v_2\dots v_n$ with 
\begin{itemize}
\item $v_\ell=w_\ell$ for any $\ell$ with $\ell\notin [a,b]$ or $w_\ell\notin [c,d]$, and
\item $red(v\,|\,([a,b],[c,d])=u$, and $v\,|\,([a,b],[c,d])$ and $w\,|\,([a,b],[c,d])$ have the same underlying alphabet.
\end{itemize}
Indeed, $v$ is obtained from $w$ by replacing the subword $w_{i_1}w_{i_2}\dots w_{i_k}$ of $w$
by an appropriate word  order-isomorphic with $u$. With these notations we call $v$ the 
{\it $([a,b],[c,d])$-substitution
by $u$ in $w$}. In particular, if $u=red(w\,|\,([a,b],[c,d])$, then the $([a,b],[c,d])$-substitution
by $u$ in $w$ is $w$ itself, and we have the following easy to understand fact.
\begin{fact}
If $red(w\,|\,([a,b],[c,d])$ and $u$ are two $d$-equivalent words, then so are $w$ and the $([a,b],[c,d])$-substitution by $u$ in $w$.
\label{first_fact}
\end{fact}

See Example \ref{Ex2} where the $([3,7],[1,4])$-substitution by $3321$ in $w=21\mathit{143}6\mathit{1}5441$ is 
$v=21\mathit{443}6\mathit{1}5441$ (the replaced elements are in italic and represented by $\times$ in
the corresponding diagrams).

\section{Proof of the main results}

The main result of this article is Theorem \ref{Main_Th}. Prior to its proof,
Lemmata \ref{Lem1} and \ref{Lem2} below establish some equidistribution results and the Corollary
\ref{main_cor2} of Theorems \ref{First_main_th} and \ref{Second_main_th} says that if two patterns
are an $f$-transformation of each other, then the patterns have the same popularity on any $d$-equivalence
class.


\begin{lem}
Let $\pi=\pi_1\pi_2\dots \pi_k$ and $\sigma=\sigma_1\sigma_2\dots \sigma_k$ be two $d$-equivalent 
patterns with $\pi_\ell=\sigma_\ell$ for any $\ell$, except $\sigma_i=\pi_i+1$ for some $i$.
Let also $t$ be a trace of both $\pi$ and $\sigma$ with one $\Box$ symbol and
$A$ be a $k-1$ element subset of $\{1,2,\ldots,n\}$. 
Then on any $d$-equivalence class the statistics $(t,A,\pi)$ and $(t,A,\sigma)$ have the same distribution.
\label{Lem1}
\end{lem}
\proof
\commm{
For any $d$-equivalence class we give a bijection $w\mapsto v$ with 
$(t,A,\pi)w=(t,A,\sigma)v$.

Since $\pi$ and $\sigma$ differ in position $i$ it follows 
that $t_i=\Box$. In addition, since $\pi$ and $\sigma$ are patterns,
the entry $\pi_i$ occurs at least twice in $\pi$ 
(otherwise $\sigma=\pi_1\ldots(\pi_i+1)\ldots\pi_k$ is not longer a pattern)
and so does $\sigma_i$ in $\sigma$.
Let $x$ be the symbol in $t$ playing the role of
$\pi_i$ and $y$ be that playing the role of $\sigma_i$,
and we define the 
interval $[c,d]=[x,y]$ and the interval $[a,b]$ as follows. 
If $A=\{p_1,p_2,\ldots,p_{k-1}\}\subset \{1,2,\ldots,n\}$, then
\begin{itemize}
\item if $i=1$, then $[a,b]=[1,p_1-1]$,
\item if $i=k$, then $[a,b]=[p_{k-1}+1,n]$,
\item elsewhere, $[a,b]=[p_{i-1}+1,p_i-1]$.
\end{itemize}

Now let $u$ be the word $red(w\,|\,([a,b],[c,d])$ and $u'$ be the word $(\mathrm{c}\circ\psi)(u)$, with
$\psi$ defined in relation (\ref{psi}) and  $\mathrm{c}$ the complement operation. The desired word $v$ is the $([a,b],[c,d])$-substitution
by $u'$ in $w$. Indeed, $u$ and $u'$ have the same descent set and same underlying alphabet, and thus they are 
$d$-equivalent. By Fact \ref{first_fact}
the transformation $w\mapsto v$ turns the word 
$w=w_1w_2\dots w_n$ into a $d$-equivalent 
word $v=v_1v_2\dots v_n$.
In addition, by property (b) of $\psi$, it follows that the number of the largest 
entries in $u$ is the same as that of the smallest entries in $u'=(\mathrm{c}\circ\psi)(u)$, and vice versa.
Thus, for any $j$, $p_{i-1}<j<p_i$ (with the convention $p_0=0$ and $p_k=n+1$)
$w\mapsto v$ transforms any occurrence 
$$
w_{p_1}\dots w_{p_{i-1}}w_jw_{p_i}\dots w_{p_{k-1}}=
t_1\dots t_{i-1}xt_{i+1}\dots t_k
$$
of $\pi$ in $w$ with trace $t$ at $A$ into an occurrence 
$$
v_{p_1}\dots v_{p_{i-1}}v_jv_{p_i}\dots v_{p_{k-1}}=
t_1\dots t_{i-1}yt_{i+1}\dots t_k,
$$
of $\sigma$ in $v$ with trace $t$ at $A$.
This transformation is reversible, indeed the $([a,b],[c,d])$-substitution by $u$ in $v$ gives the word $w$,
and so it is a bijection.
}
\endproof

\begin{exam}
We represent words as diagrams identifying words $w=w_1w_2\dots w_n$ with the set of points 
$\{(i,w_i)\,:\ 1\leq i\leq n\}$.
Let $w$ be the word $21143615441$ (the left-hand side diagram in this representation),
$\pi=1132$ and $\sigma=1232$ be two patterns as in Lemma~\ref{Lem1}, 
$t=1\Box 54$ be a common trace of $\pi$ and $\sigma$, and $A$ be the set $\{2,8,10\}$.
In the diagram representation of $w$, the entries $1$, $5$ and $4$ of $t$ occurring 
in positions belonging to $A$ are represented by \rule{0.2cm}{0.2cm} symbol.
Following the notations in the proof of 
Lemma~\ref{Lem1}, the interval $[c,d]$ is $[1,4]$, $[a,b]$ is $[3,7]$, 
$w\,|\,([a,b],[c,d])$ is the subword $1431$ (represented by $\times$ symbols in the left-hand side diagram),
$u=red(w\,|\,([a,b],[c,d])$ is the word $1321$, $u'=(\mathrm{c}\circ\psi)(u)$ is 
$\mathrm{c}(1123)=3321$, see the examples at the end of the Section \ref{sect2.3}. 
Finally, $v=21443615441$ in the right-hand side diagram is the image of $w$ through the bijection in the proof of Lemma~\ref{Lem1},
and we have $(t,A,\pi)w=(t,A,\sigma)v$. Indeed, $\pi$ occurs twice in $w$ with trace $t$ at $A$, 
namely in positions $2,3,8,10$ and in positions $2,7,8,10$; and so does $\sigma$ in $v$ with trace $t$ at $A$, namely in positions $2,3,8,10$ and in positions $2,4,8,10$.

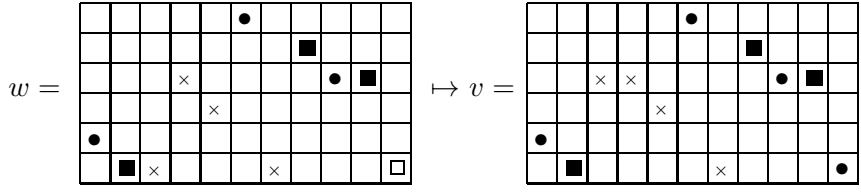
\begin{center}
\begin{tabular}{cccc}
\begin{tabular}{c}
$w=$\hspace{-2.5cm}\\
\\
\\
\\
\\
\end{tabular}
&\hspace{-0.5cm}
\unitlength=4mm
\begin{picture}(13,5)
\put(0.,0.){\line(1,0){11}}
\put(0.,1.){\line(1,0){11}}
\put(0.,2.){\line(1,0){11}}
\put(0.,3.){\line(1,0){11}}
\put(0.,4.){\line(1,0){11}}
\put(0.,5.){\line(1,0){11}}
\put(0.,6.){\line(1,0){11}}
\put(0.,0.){\line(0,1){6}}
\put(1.,0.){\line(0,1){6}}
\put(2.,0.){\line(0,1){6}}
\put(3.,0.){\line(0,1){6}}
\put(4.,0.){\line(0,1){6}}
\put(5.,0.){\line(0,1){6}}
\put(6.,0.){\line(0,1){6}}
\put(7.,0.){\line(0,1){6}}
\put(8.,0.){\line(0,1){6}}
\put(9.,0.){\line(0,1){6}}
\put(10.,0.){\line(0,1){6}}
\put(11.,0.){\line(0,1){6}}


\put(0.23,1.23){$\bullet$}
\put(1.3,0.25){\rule{0.2cm}{0.2cm}}

\put(2.2,0.3){$\scriptscriptstyle\times$}
\put(3.2,3.3){$\scriptscriptstyle\times$}
\put(4.2,2.3){$\scriptscriptstyle\times$}
\put(6.2,0.3){$\scriptscriptstyle\times$}
\put(5.23,5.23){$\bullet$}

\put(7.3,4.25){\rule{0.2cm}{0.2cm}}
\put(8.23,3.23){$\bullet$}
\put(9.3,3.25){\rule{0.2cm}{0.2cm}}
\put(10.33,0.33){\framebox(0.4,0.4){}}
\end{picture}
&
\hspace{-1.3cm}
\begin{tabular}{c}
$\mapsto$ $v=$\\
\\
\\
\\
\\
\end{tabular}
& \hspace{-0.6cm}
\unitlength=4mm
\begin{picture}(13,4)
\put(0.,0.){\line(1,0){11}}
\put(0.,1.){\line(1,0){11}}
\put(0.,2.){\line(1,0){11}}
\put(0.,3.){\line(1,0){11}}
\put(0.,4.){\line(1,0){11}}
\put(0.,5.){\line(1,0){11}}
\put(0.,6.){\line(1,0){11}}
\put(0.,0.){\line(0,1){6}}
\put(1.,0.){\line(0,1){6}}
\put(2.,0.){\line(0,1){6}}
\put(3.,0.){\line(0,1){6}}
\put(4.,0.){\line(0,1){6}}
\put(5.,0.){\line(0,1){6}}
\put(6.,0.){\line(0,1){6}}
\put(7.,0.){\line(0,1){6}}
\put(8.,0.){\line(0,1){6}}
\put(9.,0.){\line(0,1){6}}
\put(10.,0.){\line(0,1){6}}
\put(11.,0.){\line(0,1){6}}

\put(0.23,1.23){$\bullet$}
\put(1.3,0.25){\rule{0.2cm}{0.2cm}}
\put(2.2,3.3){$\scriptscriptstyle\times$}
\put(3.2,3.3){$\scriptscriptstyle\times$}
\put(4.2,2.3){$\scriptscriptstyle\times$}
\put(6.2,0.3){$\scriptscriptstyle\times$}
\put(5.23,5.23){$\bullet$}

\put(7.3,4.25){\rule{0.2cm}{0.2cm}}
\put(8.23,3.23){$\bullet$}
\put(9.3,3.25){\rule{0.2cm}{0.2cm}}
\put(10.23,0.23){$\bullet$}
\end{picture}
\end{tabular}
\end{center}
\label{Ex2}
\end{exam}

The next lemma is the counterpart of Lemma \ref{Lem1} where the patterns differ in two positions,
with the additional requirement that the two different entries occur once in each pattern.

\begin{lem}
Let $\pi=\pi_1\pi_2\dots \pi_k$ and $\sigma=\sigma_1\sigma_2\dots \sigma_k$ be two $d$-equivalent 
patterns such that there are $i$ and $j$, $i<j$, with 
\begin{itemize}
\item $\pi_\ell=\sigma_\ell$ for any $\ell$, except $\pi_i=\sigma_j$ and $\pi_j=\sigma_i$,
\item $\pi_j=\pi_i+1$,
\item each of $\pi_i$ and $\pi_j$ occurs once in $\pi$
      (or, equivalently, $\sigma_i$ and $\sigma_j$ occur once in $\sigma$).
\end{itemize}
Let also $t=t_1t_2\ldots t_k$ be a trace of both $\pi$ and $\sigma$ with two $\Box$ symbols and 
$A$ be a subset of $\{1,2,\ldots,n\}$ of cardinality $k-2$.
Then on any $d$-equivalence class the statistics $(t,A,\pi)$ and $(t,A,\sigma)$ have the same distribution.
\label{Lem2}
\end{lem}
\proof
\commm{
To a certain extent the proof is similar to that of Lemma \ref{Lem1} by giving a bijection $w\mapsto v$ with 
$(t,A,\pi)w=(t,A,\sigma)v$ on any $d$-equivalence class.\\
Since $\pi$ and $\sigma$ differ in positions $i$ and $j$ and $t$ contains two $\Box$ symbols, it follows 
that $t_i=t_j=\Box$, and since $\pi$ and $\sigma$ are $d$-equivalent 
$i$ and $j$ are not consecutive positions in $t$.
We define three intervals $[a,b]$,  $[a',b']$ and  $[c,d]$. Let 
 $A=\{p_1,p_2,\ldots,p_{k-2}\}$ be the $k-2$ element subset.
\begin{itemize}
\item If $\pi_i$ is the smallest entry in $\pi$ then $c=1$. Otherwise let 
$\pi_u$ be the largest entry in $\pi$ smaller than $\pi_i$, and $x$ be the entry in $t$
playing the role of $\pi_u$, and finally $c=x+1$. 
Similarly, if $\pi_j$ is the largest entry in $\pi$ then $d=n$. Otherwise let 
$\pi_u$ be the smaller entry in $\pi$ larger than $\pi_j$, and $x$ be the entry in $t$
playing the role of $\pi_u$, and finally $d=x-1$.
\item If $i=1$, then $a=1$, otherwise $a=p_{i-1}+1$; and $b=p_i-1$.
\item $a'=p_{j-2}+1$; and if $j=k-2$, then $b'=n$, otherwise $b'=p_{j-1}-1$.
\end{itemize}
Now we define the announced bijection $w\mapsto v$, where $v$ is obtained by constructing
the words $w'$, $w''$ and $w'''=v$ by applying the following steps.

\begin{enumerate}
\item Let $u$ be the word $red(w\,|\,([a,b],[c,d])$ and $u'=\psi(u)$, with
$\psi$ defined in relation (\ref{psi}), and $w'$ be the $([a,b],[c,d])$-substitution
by $u'$ in $w$;
\item let $u$ be the word $red(w'\,|\,([a',b'],[c,d])$ and $u'=\psi(u)$, and $w''$ be the 
$([a',b'],[c,d])$-substitution by $u'$ in $w'$;
\item let $u$ be the word $red(w''\,|\,([a,b'],[c,d])$ and $u'=\mathrm{c}(u)$,
 with $\mathrm{c}$ the complement operation, and $w'''$ be the $([a,b'],[c,d])$-substitution
by $u'$ in $w''$;
\end{enumerate}
and finally $v=w'''$.
Note that the first two steps can be performed in arbitrary order
since the substitution operations act on different entries of $w$
(on the disjoint intervals $[a,b]$ and $[a',b']$).
As in the proof of Lemma \ref{Lem1}, taking in consideration the properties of $\psi$,
$w\mapsto v$ transforms any occurrence 
of $\pi$ in $w$ with trace $t$ at $A$ into an occurrence 
of $\sigma$ in $v$ with trace $t$ at $A$, and $w\mapsto v$ is reversible
and so it is a bijection.
}
\endproof
Note that in the previous proof, unlike in that of Lemma \ref{Lem2},
the property (b) of the bijection $\psi$ is not used.

See Table \ref{Appendix_T1} in Appendix for an example of the equidistribution stated in 
Lemma \ref{Lem2}.

\begin{exam}

Let $w$ be the word $217 34 9 64 881 53 71$ (the left-hand side diagram below), 
   $\pi=125134$ and
$\sigma=135124$ be two patterns as in Lemma \ref{Lem2}, $t=1\Box 8 1 \Box 7$
be a common trace of $\pi$ and $\sigma$, and $A$ be the set $\{2,9,11,14\}$. In the diagram representation of $w$, the entries $1,8,1,7$ of $t$ occurring 
in positions belonging to $A$ are represented by \rule{0.2cm}{0.2cm} symbols.
Following the notations in the proof of 
Lemma~\ref{Lem2}, the interval $[a,b]$ is $[4,8]$, $[a',b']$ is $[12,13]$, $[c,d]$ is $[3,6]$, 
$w\,|\,([a,b],[c,d])$ is the subword $3464$ (represented by $\times$ symbols in the left-hand side diagram),
$w\,|\,([a',b'],[c,d])$ is the subword $53$ (represented by $+$ symbols),
$red(w\,|\,([a,b],[c,d])=red(3464)$ is $1232$ and $\psi(1232)=2213$, 
$red(w\,|\,([a',b'],[c,d])=red(53)$ is $21$ and $\psi(21)=12$.
Finally, $v=217 55 9 63 881 64 71$ in the left-hand side diagram is the image of $w$ through the bijection in the proof of Lemma~\ref{Lem2}, 
and we have $(t,A,\pi)w=(t,A,\sigma)v$. 
Indeed, $\pi$ occurs three times in $w$ with trace $t$ at $A$, 
namely in positions $2,4,9,11,12,14$, in positions $2,5,9,11,12,14$, and in positions
$2,8,9,11,12,14$; and so does $\sigma$ in $v$ with trace $t$ at $A$, namely in positions
$2,4,9,11,13,14$, in positions 
$2,5,9,11,13,14$, and in positions 
$2,7,9,11,13,14$.

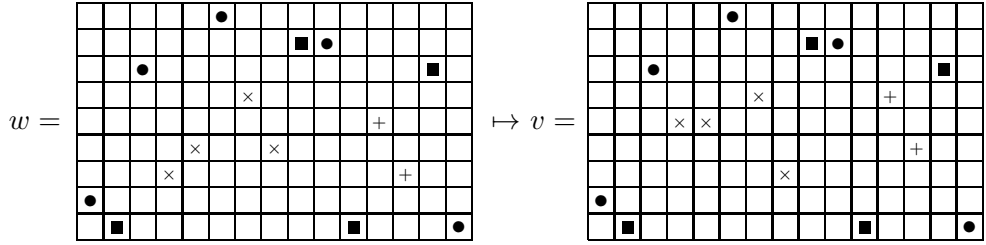
\begin{picture}(15,10)
\end{picture}
\begin{center}
\begin{tabular}{cccc}
\begin{tabular}{c}
\hspace{-1cm}$w=$\\
\\
\\
\\
\\
\\
\end{tabular}

&
\hspace{-0.7cm}
\unitlength=3.5mm
\begin{picture}(15,10)
\put(0.,0.){\line(1,0){15}}
\put(0.,1.){\line(1,0){15}}
\put(0.,2.){\line(1,0){15}}
\put(0.,3.){\line(1,0){15}}
\put(0.,4.){\line(1,0){15}}
\put(0.,5.){\line(1,0){15}}
\put(0.,6.){\line(1,0){15}}
\put(0.,7.){\line(1,0){15}}
\put(0.,8.){\line(1,0){15}}
\put(0.,9.){\line(1,0){15}}
\put(0.,0.){\line(0,1){9}}
\put(1.,0.){\line(0,1){9}}
\put(2.,0.){\line(0,1){9}}
\put(3.,0.){\line(0,1){9}}
\put(4.,0.){\line(0,1){9}}
\put(5.,0.){\line(0,1){9}}
\put(6.,0.){\line(0,1){9}}
\put(7.,0.){\line(0,1){9}}
\put(8.,0.){\line(0,1){9}}
\put(9.,0.){\line(0,1){9}}
\put(10.,0.){\line(0,1){9}}
\put(11.,0.){\line(0,1){9}}
\put(12.,0.){\line(0,1){9}}
\put(13.,0.){\line(0,1){9}}
\put(14.,0.){\line(0,1){9}}
\put(15.,0.){\line(0,1){9}}

%
\put(0.22,1.20){$\bullet$}
\put(1.3,0.25){\rule{0.15cm}{0.15cm}}
\put(2.22,6.20){$\bullet$}
\put(3.2,2.3){$\scriptscriptstyle\times$}
\put(4.2,3.3){$\scriptscriptstyle \times$}
\put(5.22,8.20){$\bullet$}
\put(6.2,5.3){$\scriptscriptstyle \times$}
\put(7.2,3.3){$\scriptscriptstyle\times$}
\put(8.3,7.25){\rule{0.15cm}{0.15cm}}
\put(9.22,7.20){$\bullet$}
\put(10.3,0.25){\rule{0.15cm}{0.15cm}}
\put(11.2,4.3){$\scriptscriptstyle +$}
\put(12.2,2.3){$\scriptscriptstyle +$}
\put(13.3,6.25){\rule{0.15cm}{0.15cm}}
\put(14.22,0.20){$\bullet$}
\end{picture}
&\hspace{-0.5cm}
\begin{tabular}{c}
$\mapsto$ $v=$\\
\\
\\
\\
\\
\\
\end{tabular}
&\hspace{-0.6cm}
\unitlength=3.5mm
\begin{picture}(15,10)
\put(0.,0.){\line(1,0){15}}
\put(0.,1.){\line(1,0){15}}
\put(0.,2.){\line(1,0){15}}
\put(0.,3.){\line(1,0){15}}
\put(0.,4.){\line(1,0){15}}
\put(0.,5.){\line(1,0){15}}
\put(0.,6.){\line(1,0){15}}
\put(0.,7.){\line(1,0){15}}
\put(0.,8.){\line(1,0){15}}
\put(0.,9.){\line(1,0){15}}
\put(0.,0.){\line(0,1){9}}
\put(1.,0.){\line(0,1){9}}
\put(2.,0.){\line(0,1){9}}
\put(3.,0.){\line(0,1){9}}
\put(4.,0.){\line(0,1){9}}
\put(5.,0.){\line(0,1){9}}
\put(6.,0.){\line(0,1){9}}
\put(7.,0.){\line(0,1){9}}
\put(8.,0.){\line(0,1){9}}
\put(9.,0.){\line(0,1){9}}
\put(10.,0.){\line(0,1){9}}
\put(11.,0.){\line(0,1){9}}
\put(12.,0.){\line(0,1){9}}
\put(13.,0.){\line(0,1){9}}
\put(14.,0.){\line(0,1){9}}
\put(15.,0.){\line(0,1){9}}
%




\put(0.22,1.20){$\bullet$}
\put(1.3,0.25){\rule{0.15cm}{0.15cm}}
\put(2.22,6.20){$\bullet$}
\put(3.2,4.3){$\scriptscriptstyle\times$}
\put(4.2,4.3){$\scriptscriptstyle \times$}
\put(5.22,8.20){$\bullet$}
\put(6.2,5.3){$\scriptscriptstyle \times$}
\put(7.2,2.3){$\scriptscriptstyle\times$}
\put(8.3,7.25){\rule{0.15cm}{0.15cm}}
\put(9.22,7.20){$\bullet$}
\put(10.3,0.25){\rule{0.15cm}{0.15cm}}
\put(11.2,5.3){$\scriptscriptstyle +$}
\put(12.2,3.3){$\scriptscriptstyle +$}
\put(13.3,6.25){\rule{0.15cm}{0.15cm}}
\put(14.22,0.20){$\bullet$}
\end{picture}
\end{tabular}
\end{center}
\end{exam}

\begin{thm}
Let $\pi=\pi_1\pi_2\dots \pi_k$ and $\sigma=\sigma_1\sigma_2\dots \sigma_k$ be two $d$-equivalent 
patterns with $\pi_\ell=\sigma_\ell$ for any $\ell$, except $\sigma_i=\pi_i+1$ for some $i$.
Then $\pi$ and $\sigma$ have the same popularity on any $d$-equivalence class.
\label{First_main_th}
\end{thm}
\proof
\commm{
By Lemma \ref{Lem1}, for any 
\begin{itemize}
\item integer $p$,
\item trace $t$ with one $\Box$ symbol in position $i$ of both $\pi$ and $\sigma$, and
\item cardinality $k-1$ subset $A$ of $\{1,2,\ldots,n\}$,
\end{itemize}
on any $d$-equivalence class
we have
$$
|\{w:(t,A,\pi)w=p\}|= |\{w:(t,A,\sigma)w=p\}|.
$$
For $t$ and $A$ fixed, summing over all $w$ in a $d$-equivalence class we have
$$
\sum_w (t,A,\pi)w=\sum_w (t,A,\sigma)w.
$$
Further, for a fixed $A$, summing over all possible traces $t$ at $A$ of both $\pi$ and $\sigma$ we have
$$
\sum_t\sum_w (t,A,\pi)w=\sum_t\sum_w (t,A,\sigma)w.
$$
Note that in this equality there are no `double counting' since different traces result in occurrences 
of $\pi$ and $\sigma$ with different 
values for the entries. Finally summing over all cardinality $k-1$ set $A$ we have
$$
\sum_A\sum_t\sum_w (t,A,\pi)w=\sum_A\sum_t\sum_w (t,A,\sigma)w.
$$
Again, there are no `double counting' since different sets $A$ result in occurrences of
$\pi$ and $\sigma$ in different positions.
The two sides of the last equality give precisely the popularity of $\pi$ and $\sigma$, respectively, 
on a $d$-equivalence class, and the statement follows.
}
\endproof

\begin{thm}
Let $\pi=\pi_1\pi_2\dots \pi_k$ and $\sigma=\sigma_1\sigma_2\dots \sigma_k$ be two $d$-equivalent 
patterns with $\pi_\ell=\sigma_\ell$ for any $\ell$, except $\pi_i=\sigma_j$ and $\pi_j=\sigma_i$
for some $i$ and $j$, and $\pi_j=\pi_i+1$.
Then $\pi$ and $\sigma$ have the same popularity on any $d$-equivalence class.
\label{Second_main_th}
\end{thm}
\proof
\commm{
We distinguish two cases: at least one of the symbols $\pi_i$ and $\pi_j$ occurs twice in $\pi$, or 
each of these symbols occurs exactly once in $\pi$.\\
In the first case, suppose that $\pi_i$ occurs twice in $\pi$ and let $\tau=\tau_1\tau_2\dots \tau_k$ be the
pattern with $\tau_\ell=\tau_\ell$ for any $\ell$, except $\tau_i=\pi_i+1$ ($=\pi_j=\sigma_i$).
Since $\pi$ and $\sigma$ are $d$-equivalent so are $\pi$ and $\tau$ (and thus $\tau$ and $\sigma$). 
By Theorem \ref{First_main_th} it follows that $\tau$ has the same popularity as $\pi$.
But $\tau_\ell=\sigma_\ell$
for any $\ell$, except $\tau_j=\sigma_j+1$ ($=\pi_j$) and again by Theorem \ref{First_main_th}
it follows that $\tau$ has the same popularity as $\sigma$, and the statement follows.\\
In the second case ($\pi_i$ and $\pi_j$ occur once in $\pi$), applying Lemma \ref{Lem2} and
reasoning as in the proof of Theorem \ref{First_main_th}, we have the desired equipopularity.
}
\endproof

Recall that an $f$-transformation turns a pattern $\pi$ into another $d$-equivalent one $\sigma$ by making  
`small changes' as in Theorems  \ref{First_main_th} and \ref{Second_main_th}, and we have the next consequence
of these theorems.

\begin{cor}
If the pattern $\sigma$  is an $f$-transformation of the pattern $\pi$, then $\pi$ and $\sigma$ have the same
popularity on any $d$-equivalence class.
\label{main_cor2}
\end{cor} 

See Table \ref{Appendix_T2} in Appendix for an example of the equipopularity stated in 
Corollary \ref{main_cor2}.
Combining Corollaries \ref{main_cor1} and \ref{main_cor2} we obtain the next theorem.

\begin{thm}
Two patterns are $d$-equivalent if and only if they have the same popularity on any 
$d$-equivalence class.
\label{Main_Th}
\end{thm}
\proof
\commm{
`$\Rightarrow$'
If the patterns $\pi$ and $\sigma$ are $d$-equivalent, then they are $f$-equivalent, and thus there is a sequence of 
patterns $\pi=\tau^{(1)},\tau^{(2)},\dots,\tau^{(p)}=\sigma$ such that $\tau^{(\ell+1)}$ is an $f$-transformation
of $\tau^{(\ell)}$, $1\leq \ell\leq p-1$. Thus $\tau^{(\ell+1)}$ and $\tau^{(\ell)}$ are equipopular 
on any $d$-equivalence class and so are $\pi$ and $\sigma$.\\
\noindent
`$\Leftarrow$'
By contraposition: if the patterns $\pi$ and $\sigma$ are not $d$-equivalent,
then within the words of the $d$-equivalence class containing (once) $\pi$
the pattern $\sigma$ does not occur, or vice versa.
Indeed $\pi$ and $\sigma$ differ by their length and/or their arity, 
and/or their descent set; and so there is a $d$-equivalence class on which 
$\pi$ and $\sigma$ are not equipopular. 
}
\endproof

Two same length words are {\it descent-equivalent} if they have same descent set
(and not necessarily same underlying alphabet),
and so $d$-equivalence implies descent-equivalence. 
A $q$-ary descent-equivalence class
is a maximal set of same length descent-equivalent $q$-ary words,
for instance $\{121,131,132,$
$231,232\}$ is a $3$-ary descent-equivalence class.
And we have the next easy to see corollary.

\begin{cor}
Two patterns are $d$-equivalent if and only if they have the same popularity on any 
$q$-ary descent-equivalence class.
\end{cor}

Finally, permutations are particular words (and particular patterns) for which 
the notions of $d$-equivalence and descent-equivalence coincide.
Specializing the previous results to permutations we have the following straightforward result.

\begin{cor}
Two permutations are descent-equivalent if and only if they have the same popularity on any 
descent-equivalence class of permutations.
\end{cor}
See Table \ref{Appendix_T2} in Appendix for an example of equipopularity
of two descent-equivalent permutations.

\bibliographystyle{plain}

\section*{Appendix}

Here we give two examples of equidistribution and equipopularity considered 
through this article.
\begin{table}[h]
\begin{center}
\begin{tabular}{ |c|c|c| }\hline  
$w$                      & $(t,A,\pi)w$               & $(t,A,\sigma)w$ \\ 
\hline
$1 5 \mathbf{4} 1 5 \mathbf{4} 3 2$&         2&       0\\
$1 5 \mathbf{4} 2 5 \mathbf{4} 3 1$&         1&       0\\
$1 5 \mathbf{4} 2 5 \mathbf{4} 3 2$&         2&       0\\
$1 5 \mathbf{4} 3 5 \mathbf{4} 2 1$&         1&       0\\
$1 5 \mathbf{4} 3 5 \mathbf{4} 3 2$&         2&       0\\
$2 5 \mathbf{4} 1 5 \mathbf{4} 3 1$&         1&       1\\
$2 5 \mathbf{4} 1 5 \mathbf{4} 3 2$&         1&       0\\
$2 5 \mathbf{4} 2 5 \mathbf{4} 3 1$&         1&       1\\
$2 5 \mathbf{4} 3 5 \mathbf{4} 2 1$&         0&       1\\
$2 5 \mathbf{4} 3 5 \mathbf{4} 3 1$&         1&       1\\
$3 5 \mathbf{4} 1 5 \mathbf{4} 2 1$&         0&       2\\
$3 5 \mathbf{4} 1 5 \mathbf{4} 3 2$&         0&       1\\
$3 5 \mathbf{4} 2 5 \mathbf{4} 2 1$&         0&       2\\
$3 5 \mathbf{4} 2 5 \mathbf{4} 3 1$&         0&       1\\
$3 5 \mathbf{4} 3 5 \mathbf{4} 2 1$&         0&       2\\
\dots & 0 & 0\\\hline
\end{tabular}
\caption{
The equidistribution of the statistics $(t,A,\pi)$ and $(t,A,\sigma)$ over
the set of length eight words with underlying alphabet $\{1,2,\dots,5\}$ and 
descent set $\{2,3,5,6,7\}$, 
for: $\pi=1332$, $\sigma=2331$, $t=\Box \mathbf{44} \Box$ and $A=\{3,6\}$.
Only words $w$ with $(t,A,\pi)w\neq 0$ or $(t,A,\sigma)w\neq 0$ are shown.
The occurrences of the symbols $\mathbf{4}$ in positions belonging to $A$ (and playing the role of $3$ in the 
occurrences of $\pi$ and of $\sigma$) are in bold.
There are six words $w$ for which $(t,A,\pi)w=1$, as many as for which $(t,A,\sigma)w=1$;
and there are theree  words $w$ for which $(t,A,\pi)w=2$, as many as for which $(t,A,\sigma)w=2$.
\label{Appendix_T1}
}
\end{center}
\end{table}

\begin{table}[h]
\begin{center}
\begin{tabular}{ |c|c|c| }\hline
$w$ & $(213)w$ & $(312)w$ \\
\hline
$2 1 3 5 4$  &     $3$  &     $0$\\
$2 1 4 5 3$  &     $3$  &     $0$\\
$3 1 2 5 4$  &     $4$  &     $1$\\
$3 1 4 5 2$  &     $2$  &     $1$\\
$3 2 4 5 1$  &     $2$  &     $0$\\
$4 1 2 5 3$  &     $2$  &     $3$\\
$4 1 3 5 2$  &     $2$  &     $2$\\
$4 2 3 5 1$  &     $2$  &     $1$\\
$5 1 2 4 3$  &     $0$  &     $5$\\
$5 1 3 4 2$  &     $0$  &     $4$\\
$5 2 3 4 1$  &     $0$  &     $3$\\
\dots        &     $0$  &     $0$\\
\hline
popularity   & $20$     &  $20$\\\hline
\end{tabular}
\end{center}
\caption{
The equipopularity of the patterns $213$ and $312$ over
the set of length five words with underlying alphabet 
$\{1,2,\dots,5\}$ (that is, length five permutations) and descent set $\{1,4\}$.
Only words $w$ with $(213)w\neq 0$ or $(312)w\neq 0$ are shown.
The two patterns are not equidistributed over this set.
\label{Appendix_T2}
}
\end{table}

\end{document}